\documentclass[12pt]{article}

\advance \textheight by \topskip
\makeatletter
\def\bsymb#1{
  \begingroup
  \let\@nomath\@gobble \boldmath
  \mathchoice
    {\hbox{$\m@th\displaystyle#1$}}
    {\hbox{$\m@th\textstyle#1$}}
    {\hbox{$\m@th\scriptstyle#1$}}
    {\hbox{$\m@th\scriptscriptstyle#1$}}
  \endgroup}
\makeatother

\def\cc#1{\kern .7em\hfill #1 \hfill\kern .7em}  
\def\ZZ{\hbox{\it Z\hskip -4.pt Z}}

\newcommand{\nc}{\newcommand}
\nc{\beq}{\begin{equation}}
\nc{\eeq}{\end{equation}}
\nc{\beqa}{\begin{eqnarray}}
\nc{\eeqa}{\end{eqnarray}}
\nc{\noi}{\noindent}

\begin{document}

\title{
\begin{flushright}
{\small USACH-FM-00/05}\\[2cm]
\end{flushright}
{\bf  On the nature of fermion-monopole supersymmetry}}

\author{{\sf 
Mikhail S. Plyushchay}\thanks{E-mail: mplyushc@lauca.usach.cl}\\
{\small {\it Departamento de F\'{\i}sica, 
Universidad de Santiago de Chile,
Casilla 307, Santiago 2, Chile}}\\
{\small 
{\it and}}\\
{\small {\it Institute for High Energy Physics, Protvino,
Russia}}}
\date{}

\maketitle
\vskip-1.0cm

\begin{abstract}
It is shown that the generator of the nonstandard fermion-monopole 
supersymmetry uncovered by De Jonghe, Macfarlane, Peeters and van Holten,
and the generator of its standard $N=1/2$ supersymmetry have to be 
supplemented by their product operator to be treated as independent supercharge. 
As a result, the fermion-monopole system possesses the nonlinear $N=3/2$ 
supersymmetry having the nature of the 3D spin-1/2 free particle's 
supersymmetry generated by the supercharges represented in a scalar form.
Analyzing the supercharges' structure, we trace how under reduction of the 
fermion-monopole system to the spherical geometry the nonlinear $N=3/2$ 
superalgebra comprising the Hamiltonian and the total angular momentum as 
even generators is transformed into the standard linear $N=1$ superalgebra 
with the Hamiltonian to be the unique even generator.
\end{abstract}

\newpage

{\bf 1.}
Some time ago, De Jonghe, Macfarlane, Peeters and van Holten
uncovered \cite{hid} that in addition to the standard 
$N=1/2$ supersymmetry \cite{hv},
the fermion-monopole system has 
the hidden supersymmetry of the nonstandard form.
The latter is characterized by the supercharge 
anticommuting for the nonlinear operator different from
the Hamiltonian and equal to the (shifted for the constant)
square of the total angular momentum operator.
The nonlinear supersymmetry of the similar form
was observed also under investigation of the space-time symmetries
in terms of the motion of pseudoclassical spinning point
particles \cite{gib,vhol,tan},
and was revealed in the 3D $P,T$-invariant systems
of relativistic fermions \cite{gps} and Chern-Simons fields \cite{np}.
In this letter we show that the generators
of the nonstandard \cite{hid} and standard \cite{hv} supersymmetries
have to be supplemented by their product operator to be treated as 
independent supercharge. As a result, 
the fermion-monopole system possesses the nonlinear $N=3/2$ 
supersymmetry having the nature of the 3D spin-1/2 free particle's 
supersymmetry generated by the supercharges represented in a scalar form.
So, the recent observation of the essentially free nature of the dynamics 
of the scalar charged particle in a monopole field \cite{monop} is extended 
for the fermion-monopole system in the context of supersymmetry.
Analyzing the supercharges' structure, we trace how under reduction of the 
fermion-monopole system to the spherical geometry the nonlinear $N=3/2$ 
superalgebra comprising the Hamiltonian and the total angular momentum as 
even generators is transformed into the standard linear $N=1$ superalgebra 
with the Hamiltonian to be the unique even generator.

The paper is organized as follows.
We start with the analysis of the supersymmetry of
the 3D spin-1/2 free particle, and then 
investigate the fermion-monopole supersymmetry
and discuss the reduction of the latter system
to the spherical geometry.
In conclusion we compare the structure of 
supersymmetries of the free fermion and fermion-monopole
systems with the structure of the $N=1$ supersymmetric
quantum mechanical systems \cite{susy}.
\vskip0.5cm

{\bf 2.}
Let us consider a 3D free spin-1/2 nonrelativistic 
particle given by the
classical Lagrangian 
\beq
\label{lan0}
L=\frac{1}{2} \dot{\bsymb{r}}{}^2+\frac{i}{2}\bsymb{\psi}\dot{\bsymb{\psi}}.
\eeq
The corresponding Hamiltonian is
$H=\frac{1}{2}\bsymb{p}^2$,
and nontrivial Poisson-Dirac brackets are $\{r_i,p_j\}=\delta_{ij}$ and
\beq
\label{susy0}
\{\psi_i,\psi_j\}=-i\delta_{ij}.
\eeq
The set of even integrals of motion
is given by the vectors $\bsymb{p}$,
$\bsymb{L}=
\bsymb{r}\times \bsymb{p}$ and $\bsymb{K}= \bsymb{p}\times \bsymb{L}$,
which have a nonzero projection on a unit of Grassmann algebra,
and by the nilpotent spin vector $\bsymb{S}=
-\frac{i}{2}\bsymb{\psi}\times\bsymb{\psi}$
generating rotations of odd Grassmann variables $\psi_i$.
The vector $\bsymb{K}$ is the analog of the Laplace-Runge-Lentz
vector of the Kepler system, which together with
$\bsymb{p}$ and $\bsymb{L}$ constitute a  
non-normalized basis of orthogonal vectors
forming a nonlinear algebra with nontrivial part
given by the relations
\beq
\label{so31}
\{L_i,V_j\}=\epsilon_{ijk}V_k,\quad V_i=p_i,L_i,K_i,\quad
\{K_i,p_j\}=\bsymb{p}^2 \delta_{ij}-\bsymb{p}_i\bsymb{p}_j,\quad
\{K_i,K_j\}=-\bsymb{p}^2\epsilon_{ijk}L_k.
\eeq
The variables $\psi_i$ form the odd vector integral of motion,
and their algebra (\ref{susy0}) is the classical analog of Clifford algebra
with three generators.
Projecting the odd vector $\bsymb{\psi}$
onto the even vector integrals of motion,
we get three odd scalar integrals of motion (supercharges)
\beq
\label{susysc}
Q_1=\bsymb{p\psi},\quad
Q'_2=\bsymb{L\psi},\quad
Q_3=\bsymb{K\psi}.
\eeq
The supercharge $Q_1$ is a ``square root from the Hamiltonian",
$\{Q_1,Q_1\}=-2iH$.
It has zero bracket
with the supercharge $Q_3$, $\{Q_1,Q_3\}=0$,
but $Q_1$ and $Q_3$ have nontrivial brackets with $Q'_2$,
in particular, $\{Q_1,Q'_2\}=2i\bsymb{LS}$.
One can find the linear combination of 
the odd scalar integrals $Q'_2$ and $i\epsilon_{ijk}\psi_i\psi_j\psi_k$
having zero brackets with other two supercharges,
\beq
\label{q2}
Q_2=\bsymb{L\psi}-\frac{i}{3}(\bsymb{\psi}\times\bsymb{\psi})\cdot\bsymb{\psi}.
\eeq
Finally, we get the set of three scalar supercharges, $\{J_i,Q_a\}=0, $
$a=1,2,3$,
forming together with $H$ and $\bsymb{J}$
the nonlinear superalgebra
\beqa
&\{Q_1,Q_1\}=-2iH,\quad
\{Q_2,Q_2\}=-i\bsymb{J}^2,\quad
\{Q_3,Q_3\}=-2iH\cdot\bsymb{J}^2,&\label{qiqi}\\
&\{Q_a,Q_b\}=0, \quad a\neq b,&\label{qiqi0}
\eeqa
where $\bsymb{J}=\bsymb{L}+\bsymb{S}$ is the total angular
momentum vector.
The scalar supercharges
satisfy also the algebraic relations:
\beq
\label{sualg}
Q_1Q_2=i\tilde{Q}_3,\quad
Q_1Q_3=-2iH\cdot \tilde{Q}_2,\quad
Q_2Q_3=i\bsymb{J}^2\cdot \tilde{Q}_1.
\eeq
Here $\tilde{Q}_a$ means $Q_a$ with
odd vector $\bsymb{\psi}$ changed for even $\bsymb{S}$.
Since $\bsymb{J}^2\cdot\tilde{Q}_1=\bsymb{L}^2\cdot\tilde{Q}_1$,
relations (\ref{sualg}) reflect
the analogous relations between the even
vector integrals $\bsymb{p}$, $\bsymb{L}$ and $\bsymb{K}$:
\beq
\label{eventri}
\bsymb{p}\times \bsymb{L}=\bsymb{K},\quad
\bsymb{p}\times \bsymb{K}=-2H\cdot\bsymb{L},\quad
\bsymb{L}\times \bsymb{K}=\bsymb{L}^2\cdot \bsymb{p}.
\eeq
Taking into account the equalities  $\bsymb{J}^2=2H\cdot \Delta$ and 
\beq
\label{delt}
\{\Delta,H\}=\{\Delta,Q_a\}=0,
\eeq
where 
$\Delta=\bsymb{r}^2_\perp+(\bsymb{LS})\cdot{H}^{-1}$,
$\bsymb{r}_\perp=\bsymb{r}-\bsymb{p}(\bsymb{pr})\cdot\bsymb{p}^{-2}=
\bsymb{K}\cdot\bsymb{p}^{-2}$,
one can transform (at $\bsymb{p}^2\neq 0$, $\bsymb{K}^2\neq 0$)
the set of scalar supercharges
$Q_a$ into the set 
\beq
\label{barq}
\bar{Q}_1=Q_1,\quad
\bar{Q}_2=Q_2\cdot\Delta^{-1/2},\quad
\bar{Q}_3=Q_3\cdot(2H\Delta)^{-1/2}.
\eeq
This set of (nonlinearly) transformed scalar
supercharges gives rise to the $N=3/2$ linear superalgebra of the standard
form,
\beq
\label{susys}
\{\bar{Q}_a,\bar{Q}_b\}=-2iH\delta_{ab}.
\eeq

The quantum analogs of the odd variables can be realized
via the Pauli matrices (we use the system of units
$\hbar=c=1$), $\hat{\psi}_i=\frac{1}{\sqrt{2}}\sigma_i$,
and the quantum spin vector $\hat{\bf S}$ is 
proportional (``parallel") to $\hat{\bsymb{\psi}}$:
$\hat{\bf S}=\frac{1}{2}\bsymb{\sigma}=\frac{1}{\sqrt{2}}\hat{\bsymb{\psi}}$.
Note that classically this property is reflected
in the relation $\bsymb{\psi}\times {\bf S}=0$.
To construct the quantum analogs of the supercharges
$Q_a$ in the form of Hermitian operators,
we choose the natural prescription ${\bf p}\times {\bf L}\rightarrow
\frac{1}{2}(\hat{\bf p}\times\hat{\bf L}+(\hat{\bf p}\times\hat{\bf L})^\dagger)$,
and get 
\beq
\label{quq0}
\hat{Q}_1=\hat{\bf p}\hat{\bsymb{\psi}},\quad
\hat{Q}_2=\hat{\bf L}\hat{\bsymb{\psi}}+\frac{1}{\sqrt{2}},\quad
\hat{Q}_3=(\hat{\bf p}\times\hat{\bf L})\hat{\bsymb{\psi}}-i\hat{Q}_1.
\eeq
These operators are Hermitian and 
satisfy the quantum relations (there is no summation 
in repeated indexes)
\beqa
&\hat{Q}_a\hat{Q}_b=\frac{1}{2}\hat{A}_a\delta_{ab}+
\frac{i}{\sqrt{2}}\hat{B}_c\epsilon_{abc}\hat{Q}_c,&
\label{qususy00}\\
&\hat{A}_1=2\hat{H},\quad
\hat{A}_2=\hat{\bf J}{}^2+\frac{1}{4},\quad
\hat{A}_3=\hat{A}_1\hat{A}_2,&\label{qususy01}\\
&\hat{B}_1=\hat{A}_2,\quad
\hat{B}_2=\hat{A}_1,\quad
\hat{B}_3=1.&\label{qusy02}
\eeqa
The symmetric part of relations (\ref{qususy00})
is the nonlinear $N=3/2$ superalgebra  being the exact quantum analog
of classical relations (\ref{qiqi}), (\ref{qiqi0}),
whereas,
with taking into account the above mentioned 
relation between $\hat{S}_i$ and $\hat{\psi}_i$, we find that
the antisymmetric part of (\ref{qususy00}) corresponds to the 
classical relations (\ref{sualg}).

Due to the relation $\hat{Q}_3=-i\sqrt{2}\hat{Q}_1\hat{Q}_2$,
it seems that one could interpret $\hat{Q}_1$ and $\hat{Q}_2$
as the ``primary" supercharge operators and $\hat{Q}_3$ as the 
``secondary" operator. But such an interpretation
is not correct. Indeed, 
on the one hand, the relations (\ref{qususy00})
can be treated as a quantum generalization of the
symmetric in indexes relations for the three Clifford algebra
generators $\sigma_i=\sqrt{2}\hat{\psi}_i$, 
$\sigma_i\sigma_j=\delta_{ij}+i\epsilon_{ijk}\sigma_k$,
and we remember that the operators $\hat{\psi}_i$
are the quantum analogs of the set of
the classical odd integrals of motion $\psi_i$ 
forming a 3D vector.
On the other hand, the antisymmetric part of (\ref{qususy00})
is a reflection of the 
cyclic classical relations (\ref{eventri}).
It is worth noting also that 
we can pass from the set of odd integrals 
(\ref{barq}) (constructed at 
$\bsymb{p}^2\neq 0$, $\bsymb{K}^2\neq0$)
to the integrals 
\beq
\label{trgr}
\xi_a=\bar{Q}_a\cdot(2H)^{-1}.
\eeq
Then, due to the relations $\{\xi_a,\xi_b\}=-i\delta_{ab}$,
one can treat the transition from the integrals $\psi_i$
to the integrals $\xi_a$ as a simple
canonical transformation for the odd 
sector of the phase space (note, however, that $\xi_a$
have nontrivial brackets with even variables).
Formally, the same unitary in the odd sector transformations
can be realized at the quantum level.
So, it is natural to treat
all the three supercharge operators $\hat{Q}_a$
on the equal footing.

\vskip0.5cm

{\bf 3.}
Let us consider the case of fermion particle with charge $e$ in 
arbitrary time-independent magnetic field $\bsymb{B}(\bsymb{r})=
\bsymb{\nabla}\times
\bsymb{A}(\bsymb{r})$. At the Hamiltonian level
this corresponds to the change of the canonical
momentum vector $\bsymb{p}$ for the vector
$\bsymb{P}=\bsymb{p}-e\bsymb{A}$,
$\{P_i,P_j\}=e\epsilon_{ijk}B_k$. 
By projecting the odd vector variable $\bsymb{\psi}$
onto $\bsymb{P}$, we get the scalar $Q_1=\bsymb{P\psi}$.
Identifying the bracket $\frac{i}{2}\{Q_1,Q_1\}$
as the Hamiltonian,
\beq
\label{hmag}
H=\frac{1}{2}\bsymb{P}^2-e\bsymb{BS},
\eeq
$Q_1$ is automatically the odd integral of motion
(supercharge).
However, now, unlike the case of a free particle,
either even vectors $\bsymb{P}$, $\bsymb{L}=\bsymb{r}\times\bsymb{P}$,
$\bsymb{P}\times\bsymb{L}$, $\bsymb{S}$
or odd vector $\bsymb{\psi}$ are not integrals of motion,
whereas the odd scalar $(\bsymb{\psi}\times\bsymb{\psi})\cdot\bsymb{\psi}$
is conserved. Having in mind the analogy with 
the free particle case, let us check
other odd scalars for their possible conservation.
First, it is worth noting that the quantum 
relation $\hat{\bsymb{S}}=\frac{1}{\sqrt{2}} 
\hat{\bsymb{\psi}}$ is reflected classically
also in the form of the identical evolution,
$\dot{\bsymb{\psi}}=e\bsymb{\psi}\times
\bsymb{B}$, $\dot{\bsymb{S}}=e\bsymb{S}\times\bsymb{B}$.
Like the projection of the odd vector
$\bsymb{\psi}$,
the projection of $\bsymb{S}$  on $\bsymb{P}$
is conserved,
i.e. as in a free case, $\tilde{Q}_1=\bsymb{PS}$ is the integral
of motion, but generally
$\frac{d}{dt}(\bsymb{SL})=e(\bsymb{S}\times\bsymb{P})\cdot
(\bsymb{r}\times\bsymb{B})\neq 0$.
The scalar $\tilde{Q}{}'_2=\bsymb{SL}$
is the integral of motion only if
$\bsymb{B}=f(r)\cdot\bsymb{r}$, $r=\sqrt{\bsymb{r}^2}$.
This corresponds to the case of the monopole field,
for which $f(r)=gr^{-3}$ ($\bsymb{r}\neq 0$, $g=const$) 
is fixed by the condition 
$\bsymb{\nabla B}=0$, and from now on, we restrict the analysis
by the fermion-monopole system.
Though in this case
the brackets $\{L_i,L_j\}=\epsilon_{ijk}(L_k+\alpha n_k)$,
$\alpha=eg$,
are different from the corresponding brackets for the free
particle, nevertheless the odd scalar $Q'_2=\bsymb{L\psi}$ 
is the integral of motion.
One can check that
the direct analog of the free particle's supercharge
(\ref{q2}) has zero bracket with $Q_1$
and that
\beq
\label{q2mon}
\{Q_2,Q_2\}=-i(\bsymb{J}^2-\alpha^2).
\eeq
Here $\bsymb{J}=\bsymb{L}+\bsymb{S}-\alpha{\bf r}\cdot r^{-1}$
is the conserved angular momentum vector
of the fermion-monopole system,
whose components form $su(2)$ algebra,
$\{J_i,J_j\}=\epsilon_{ijk}J_k$, and generate rotations.
The scalar $Q_3=(\bsymb{P}\times \bsymb{L})\bsymb{\psi}$
is also the integral of motion and
in the fermion-monopole case 
the classical relations of the form 
(\ref{qiqi}), (\ref{qiqi0}), (\ref{sualg})
take place with the change $\bsymb{J}^2\rightarrow
\bsymb{J}^2-\alpha^2$ and with the Hamiltonian
given by Eq. (\ref{hmag}). Like for the free particle,
the superalgebra can be reduced to the standard linear
form (\ref{susys}) via the nonlinear transformation (\ref{barq}),
for which in the present case we proceed from the relation 
$\bsymb{J}^2-\alpha^2=2H\cdot\Delta$
with $\Delta=\tilde{\bsymb{r}}{}^2+
\bsymb{LS}\cdot H^{-1}$, 
$\tilde{\bsymb{r}}{}^2=\bsymb{r}^2-
(\bsymb{Pr})^2\cdot(2H)^{-1}$.
The quanitity  $\tilde{\bsymb{r}}{}^2$
is the integral of motion which in the case of
the scalar charged particle ($\psi_i=0$)
in the field of monopole
gives a minimal charge-monopole distance
in the point of perihelion: 
$r_{min}=\sqrt{\tilde{\bsymb{r}}{}^2}$ \cite{monop}.

Constructing the quantum analogs of the supercharges
in the same way as in the free particle case,
we get the supercharge operators of the
form (\ref{quq0}) with 
$\hat{\bsymb{p}}$ changed for $\hat{\bsymb{P}}$.
They satisfy the set of (anti-)commutation relations
of the same form 
(\ref{qususy00})--(\ref{qusy02})
with 
$\hat{A}_2=\hat{\bsymb{J}}{}^2+\frac{1}{4}$ changed
for $\hat{\bsymb{J}}{}^2-\alpha^2+\frac{1}{4}$,
where $\alpha$ is subject to the Dirac quantization
condition, $2\alpha\in \ZZ$.

The fermion-monopole
integral $\hat{Q}_2$ was observed for the first time 
on algebraic grounds by d'Hoker and Vinet \cite{hv}
in the context of generalization of the so called
dynamical symmetries of the charge-monopole system \cite{jac}  
for the supersymmetric case. 
The supercharge nature of $\hat{Q}_2$ and
the associated nonstandard nonlinear superalgebra 
of the operators $\hat{Q}_2$ and $\hat{Q}_1$,
was uncovered by De Jonghe, Macfarlane, Peeters and van Holten
\cite{hid}.
The present analysis shows that the set of 
supercharge operators $\hat{Q}_1$ and $\hat{Q}_2$ has
to be extended by the scalar integral $\hat{Q}_3$,
and these three odd operators together with even
operators $\hat{H}$ and $\hat{\bsymb{J}}$
form the described nonlinear $N=3/2$ superalgebra.
As we have seen, this
nonlinear supersymmetry of the fermion-monopole system
has the nature of the  free fermion particle's supersymmetry
generated by the supercharges represented in a scalar form.

{}Comparing the fermion-monopole
Hamiltonian (\ref{hmag}) (with $\bsymb{B}=\alpha \bsymb{r}\cdot r^{-3}$)
with the free fermion particle Hamiltonian
$H=\frac{1}{2}\bsymb{p}^2$, it seems that they have rather
different structure, but this is not so,
and their similarity can be revealed, like in the case
of the scalar particle \cite{monop},
by separating the even phase space coordinates into the radial
and angular ones.
The radial coordinates for the fermion-monopole system are $r$
and $P_r=\bsymb{Pn}$,
and the angular phase space variables are $\bsymb{n}=\bsymb{r}\cdot r^{-1}$,
$\bsymb{\cal J}=\bsymb{L}-\alpha\bsymb{n}$.
These coordinates have the nontrivial brackets
$\{r,P_r\}=1$, $\{{\cal J}_i,{\cal J}_j\}=
\epsilon_{ijk}{\cal J}_k$,
$\{{\cal J}_i,n_j\}=\epsilon_{ijk}n_k$,
and satisfy the relations $\bsymb{\cal J}\bsymb{n}=-\alpha$,
$\bsymb{n}^2=1$. 
In terms of these variables, the Hamiltonian of the
fermion-monopole system is
$$
H=\frac{1}{2}P_r^2+\frac{\bsymb{J}^2-\alpha^2}{2r^2}
-\frac{\bsymb{LS}}{r^2},
$$
with $\bsymb{J}=\bsymb{\cal J}+\bsymb{S}$.
The case  of the free fermion corresponds to $\alpha=0$,
and its Hamiltonian takes the similar form
$
H=\frac{1}{2}p_r^2+\frac{1}{2r^2}\bsymb{J}^2
-\frac{1}{r^2}\bsymb{LS},
$
where 
$\bsymb{J}=\bsymb{L}+\bsymb{S}$.
The difference between the two systems is
encoded now in the topology of even 
angular phase space variables \cite{monop}.

\vskip0.5cm

{\bf 4.}
Let us look at the fermion-monopole supersymmetry
from the point of view
of the reduction of the system to the
spherical geometry. 
To this end we first note that the bracket of the supercharge $Q_2$ 
with itself can be represented in the form
\beq
\label{q2red}
i\{Q_2,Q_2\}=2r^2\left(H-
\frac{1}{2}P_r^2+\frac{i}{r}Q_1(\bsymb{\psi n})\right),
\eeq
and the supercharge $Q_3$ can be reduced to the equivalent form
\beq
\label{q3red}
Q_3=2Hr\left(\bsymb{\psi n}-Q_1\frac{P_r}{2H}\right).
\eeq
As it was shown in ref. \cite{hid},
the reduction of the fermion-monopole system
to the spherical geometry can be realized
by introducing into the system the classical relations 
\beqa
&\bsymb{r}^2-\rho^2=0,\quad
P_r=0,&\label{sfera1}\\
&\bsymb{\psi n}=0,&\label{sfera2}
\eeqa
which have to be treated as the set of second class constraints
with  $\rho\neq0$ being a constant, and for simplicity we fix 
it in the form $\rho=1$.
The relations (\ref{q2red}) and (\ref{q3red}) 
allow us to observe directly that 
the described $N=3/2$ nonlinear fermion-monopole
supersymmetry is transformed into  the $N=1$ supersymmetry
of the standard linear form in the case
of reduction (\ref{sfera1}), (\ref{sfera2}).
Indeed, after reducing the fermion-monopole system
onto the surface of even second class constraints (\ref{sfera1}),
we find that the structure of the supercharge $Q_3$ is 
trivialized and takes the form of 
the odd scalar $\bsymb{\psi n}$ multiplied by $2H$.
Two other supercharges $Q_1$ and $Q_2$ after
such a reduction take the form of  linear combinations 
of the odd vector $\bsymb{\psi}$ projected 
on the vectors $\bsymb{\cal J}+\alpha\bsymb{n}$ and 
$\bsymb{\cal J}\times\bsymb{n}$ orthogonal to $\bsymb{n}$.
Then taking into account the odd second class constraint
(\ref{sfera2}) results in eliminating the supercharge $Q_3$
and in reducing the bracket 
(and corresponding anticommutator at the quantum level) 
of the supercharge $Q_2$ to $2h$, 
where $h$ is the reduced Hamiltonian, 
\beq
\label{hreds}
h=\frac{1}{2}(\bsymb{J}^2-\alpha^2).
\eeq
In other words, the supersymmetry of the fermion-monopole
system in spherical geometry is
reduced to the standard linear $N=1$ supersymmetry
characterized by two supercharges anticommuting
for the Hamiltonian \cite{susy}.
More explicitly, after reduction to the surface of the
second class constraints, the radial variables
$r$ and $P_r$ are eliminated from the theory.
The even variables can be  represented by
the total angular momentum $\bsymb{J}$ and by the unit vector
$\bsymb{n}$ having the nontrivial Dirac brackets
coinciding with corresponding initial Poisson brackets,
$\{J_i,J_j\}^*=\epsilon_{ijk}J_k$,
$\{J_i,n_j\}^*=\epsilon_{ijk}n_k$.
The odd variables $\psi_i$ satisfy 
the relation (\ref{sfera2}) which has to be treated as a strong 
equality, and their nontrivial Dirac brackets 
are $\{\psi_i,\psi_j\}^*=-i(\delta_{ij}-n_in_j)$,
$\{J_i,\psi_j\}^*=\epsilon_{ijk}\psi_k$.
The even and odd variables are subject
also to the relation
$\bsymb{Jn}=-\alpha+iq_1q_2\cdot(2h)^{-1}$,
where $h$ is given by Eq. (\ref{hreds})
and 
\beq
\label{q12}
q_1=(\bsymb{J}\times\bsymb{n})\bsymb{\psi},\quad
q_2=\bsymb{J\psi}
\eeq
are the supercharges $Q_1$ and $Q_2$
reduced to the surface (\ref{sfera1}), (\ref{sfera2}).
With the listed Dirac brackets, one can easily 
check that now the reduced supercharges
(\ref{q12}) satisfy 
the superalgebra of the standard $N=1$ supersymmetry:
$\{q_\mu,q_\nu\}^*=-2i\delta_{\mu\nu}h$, 
$\{q_\mu,h\}=0$, $\mu,\nu=1,2$.

\vskip0.5cm

{\bf 5.}
To conclude, let us compare the structure of
supersymmetric quantum mechanics with the structure of the free fermion 
and fermion-monopole systems.
A supersymmetric quantum mechanical system \cite{susy}
is characterized by the Hamiltonian
$\hat{H}=\frac{1}{2}(\hat{p}{}^2+W^2(x)+\sigma_3
W'(x))$ with $\hat{p}=-id/dx$, and by the supercharges 
$\hat{Q}_1=\hat{\theta}_1W(x)-\hat{\theta}_2
\hat{p}$, $\hat{Q}_2=\hat{\theta}_2W(x)+\hat{\theta}_1\hat{p}$
with
$\hat{\theta}_\mu=\frac{1}{\sqrt{2}}\sigma_\mu$, $\mu=1,2$,
which form the $N=1$ superalgebra
$[\hat{Q}_\mu,\hat{Q}_\nu]_{{}_{+}}=2\delta_{\mu\nu}\hat{H}$, 
$[\hat{H},\hat{Q}_\mu]=0$. The operator 
$\hat{Q}_3=\frac{1}{\sqrt{2}}\sigma_3$ is 
the trivial integral of motion, $[\hat{Q}_3,\hat{H}]=0$,
and the set of supercharges $\hat{Q}_{1,2}$
together with $\hat{Q}_3$ satisfy the relations of
the form (\ref{qususy00}) with $\hat{A}_1=\hat{A}_2=
2\hat{B}_3=2\hat{H}$,
$\hat{A}_3=\hat{B}_1=\hat{B}_2=1$.
In this case the operator $\sigma_3=-2i\hat{\theta}_1\hat{\theta}_2=
\sqrt{2}\hat{Q}_3$ (being analogous to 
any component of the odd vector integral  
$\sqrt{2}\hat{\bsymb{\psi}}$
of the free fermion system)
plays the role of the grading operator
commuting (anticommuting) with operators $\hat{x}$ and $\hat{p}$
($\hat{\theta}_{1,2}$), and, as a consequence,
commuting (anticommuting) with the Hamiltonian $\hat{H}$
(supercharges $\hat{Q}_{1,2}$).
On the other hand, for the fermion-monopole and the free fermion systems,
one can construct the operator 
$\sqrt{2}\hat{\xi}_3$ proceeding from the classical
relation (\ref{trgr}). It seems that such operator 
could be treated as the grading operator due to the relation 
$2\hat{\xi}_3^2=1$ and its anticommutation with the
supercharges $\hat{Q}_{1,2}$.
But such interpretation is not correct
since unlike the case of
supersymmetric quantum
mechanics, this operator 
has a nontrivial dependence on even operators,
and as a consequence, does not commute
with them.
 
The ordinary form of the 
classical Lagrangian for the supersymmetric quantum mechanics,
\beq
\label{lsusyqm}
L=\frac{1}{2}(\dot{x}^2-W^2(x)-2iW'(x)\theta_1\theta_2
+i\theta_\mu\dot{\theta}_\mu),
\eeq
contains only two Grassmann variables.
At the classical level the even quantity $-2i\theta_1\theta_2$
corresponds to the odd operator $\sigma_3$ 
(the latter being one of three generators of the corresponding 
Clifford algebra), i.e. here we have some sort
of classical anomaly \cite{gm}.
However, the symmetry between quantum and classical
pictures can easily be restored (``the anomaly can be canceled")
extending the set $\theta_{\mu}$, $\mu=1,2$, by
the independent Grassmann variable $\theta_3$
and changing Lagrangian (\ref{lsusyqm})
for 
$
{\cal L}=L+\frac{i}{2}\theta_3\dot{\theta}_3.
$
Such a classical system 
has two nontrivial supercharges $Q_1$,
$Q_2$, and the third odd integral of motion
given by $\theta_3$ (like $\psi_i$ for the free fermion) is
trivial and completely decoupled from other variables. 
Therefore,  
the difference of the superalgebraic structures
of the supersymmetric quantum mechanics on the one hand 
and fermion-monopole system on the other hand
is also reflected in the absence in the latter case
of the grading
operator commuting with the Hamiltonian and anticommuting
with any two of three scalar supercharges and 
which simultaneously would commute with the
initial coordinate and momenta operators.
 
\vskip0.3cm
{\bf Acknowledgements}
\vskip3mm

The work has been supported in part by the
grant 1980619 from FONDECYT (Chile)
and by DICYT (USACH).

\end{document}